Integration of Quantum, Statistical, and Irreversible Thermodynamics in A Coherent Framework

Zi-Kui Liu

Department of Materials Science and Engineering, The Pennsylvania State University,

University Park, Pennsylvania 16802, USA

**Abstract:**

The combined law of thermodynamics derived by Gibbs laid the foundation of thermodynamics though only applicable to systems without internal processes.  Gibbs further derived the classical statistical thermodynamics in terms of the probability of configurations in a system, which was extended to quantum mechanics-based statistical thermodynamics by Landau, while the irreversible thermodynamics was systemized by Onsager and expanded to chemical reactions by Prigogine.  The development of density function theory (DFT) by Kohn enabled the quantitative prediction of properties of the ground-state configuration of a system from quantum mechanics. Here, we will present our theories that integrate quantum, statistical, and irreversible thermodynamics in a coherent framework by utilizing the predicative capability of DFT to revise the statistical thermodynamics (zentropy theory) and by keeping the entropy production due to irreversible processes in the combine law of thermodynamics to derive flux equations (theory of cross phenomena).  The zentropy theory is shown capable of predicting the free energy landscape including singularity and instability at critical point and emergent positive or negative divergences of properties. The theory of cross phenomena can predict the coefficients of internal processes between conjugate variables (direct phenomena) and non-conjugate variables (cross phenomena) in the combined law of thermodynamics.  Both are with inputs from DFT-based calculations only and without fitting parameters.

# 1 Introduction

In last several centuries, the accumulation of scientific knowledge has dramatically enhanced our understanding of nature. One of the most significant events is the discovery of quantum mechanics (QM) along with the numerical solutions of the QM Schrödinger equation [1,2] offered by the density functional theory (DFT) [3,4]. Formulated as an exact theory of many-body systems, DFT [3] articulates that for an interacting electron gas there exists a universal functional of the density such that the energy is at its minimum value, i.e., the ground-state energy with a unique ground-state electron density. The numerical solution is formulated by explicitly separating the independent-electron kinetic energy and long-range Coulomb interaction energy and replacing the many-body electron problem using independent valence electrons with an exchange-correlation functional of the electron density and an associated exchange-correlation energy and potential [4], i.e., *coarse graining of electrons*. DFT-based first-principles calculations have resulted in ever-growing massive digital databases of properties of matters predicted using high-performance computers with a speed far exceeding the previous accumulations of data and knowledge by observations and experimentations. Many of these databases are part of the Open Databases Integration for Materials Design (OPTIMADE) consortium [5] with a universal application programming interface (API) to make materials databases accessible and interoperable with an extensive list of database providers [6], including the Material-Property-Descriptor Database (MPDD) from the author's group [7,8].

While those databases and other DFT-based predictions have made significant strides in the scientific community, their quantitative agreements with experimental observations are lacking with the reasons discussed in the publication co-authored by the present author [9]. The central issues limiting the quantitative predictions are that the state-of-the-art approaches use effective Hamiltonian to fit underlying interactions to few selected DFT predictions and involve truncation in terms of interaction types and ranges. Furthermore, additional specialized terms in the Hamiltonian are needed to accommodate coupling between various types of degrees of freedoms. Furthermore many scientific efforts for improving DFT methods to obtain better electron density and total energy [10–16] include time-dependent DFT (TDDFT) [17–19], random phase approximation (RPA) [20,21], density-matrix functional theory (DMFT) [22–25], DFT+U [26–28], dynamical mean-field theory [29,30], benchmarking with experimental measurements [31], deep neural network machine learning models [32–34], and some other hybrid methods [35]. These approaches aim to either improve the calculations for the ground-state configuration or extend to excited non-ground-state configuration, i.e., beyond 0 K. Nevertheless, these bottom-up predictions have not resulted in fully satisfactory quantitative agreements with experiments without fitting parameters.

On the other hand, a system at finite temperature can be considered as a statistical mixture of various configurations from the top-down perspective of the system. In the statistical mechanics introduced in 1901, Gibbs [36] considered "a great number of independent systems (states) of the same nature (of a system), but differing in the configurations and velocities which they have at a given instant, and differing not merely infinitesimally, but it may be so as to embrace every conceivable combination of configuration and velocities". In formulating statistical mechanics in the QM framework, Landau and Lifshitz [37] considered quantum configurations corresponding to the energy interval equal in the order of magnitude to the mean fluctuation of energy of the system. By correlating the number of quantum states with the particle states in the limit of the classical theory, they obtained the same entropy of a system as Gibbs.

The central task is thus to define the foundational configurations of a system as its building blocks. The multiscale entropy approach [38–41] (recently termed as zentropy theory [42–45]) starts with the ground-state configuration defined by DFT and enumerates ergodic non-ground-state configurations through symmetry-breaking degrees of freedom of the ground-state configuration. The partition function of the system at finite temperatures is obtained by the sum of the partition functions of all the configurations. Since those configurations are not pure quantum configurations, their partition functions are obtained from their free energy rather than total energies in the quantum statistical mechanics. While Gibbs statistical mechanics did not explicitly specify pure quantum configurations as QM was not developed at that time, it included only entropy among configurations, but not entropy in each configuration. The zentropy theory postulates that the total entropy of the system should include both entropies in each configuration and among the configurations.

## 2 Zentropy theory: Integration of quantum mechanics and statistical mechanics

Zentropy theory postulates that for a system composed of many configurations, its entropy can be written as follows.

$$S = \sum_{k=1}^{m} p^k S^k - k_B \sum_{k=1}^{m} p^k \ln p^k \qquad \text{Eq. 1}$$

where $m$ is the number of configurations, $k_B$ is the Boltzmann constant, and $p^k$ and $S^k$ are the probability and entropy of configuration $k$, respectively. It can be seen that the first summation in Eq. 1 represents the bottom-up view of the system considering the contribution from individual configurations of the system, while the second summation denote the top-down view of the system seeing the statistical fluctuations of configurations. Consequently, each snapshot of the system is one representation of configurations and thus different from each other statistically, and sufficient sampling of the snapshots thus reproduces the macroscopic behaviors of the system. By examining the local structures in those snapshots, one may be able to obtain the foundational configurations of the system as shown experimentally [46] or computationally [47] for $PbTiO_3$. Furthermore, in principle, this nested formula can be extended to consider more complex systems such as black holes with more degrees of freedom with Eq. 1 denoting one of the configurations of the system [43,44], and in another direction to configurations with fewer degrees of freedom within the configuration $k$ until it reaches the configurations that DFT deals with, i.e., the ground-state and non-ground-state configurations [40] as schematically shown in Figure 1. The latter may provide some insights into superconducting and other interesting ground-state configurations as postulated by the present author [43,44].

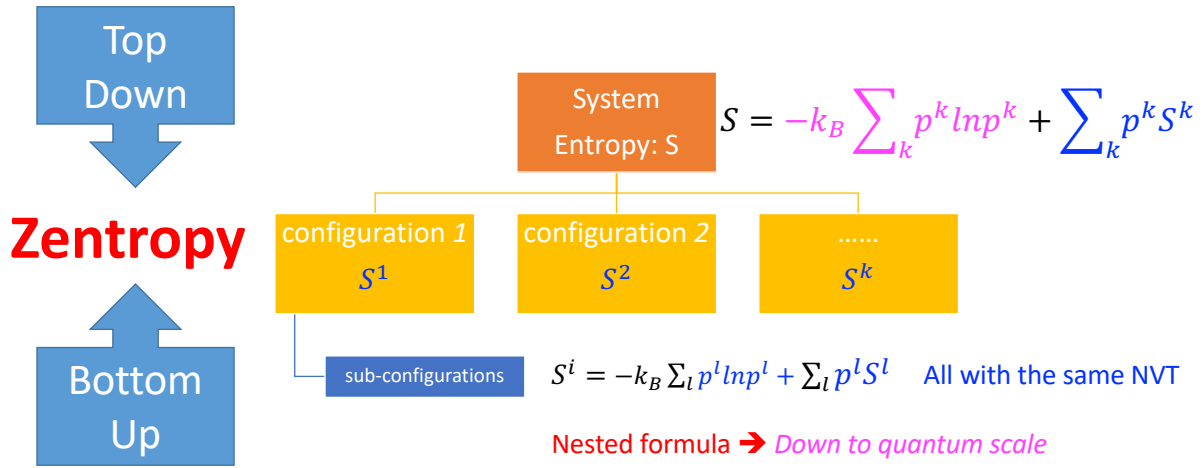

Figure 1: Schematic top-down and bottom-up integration of the zentropy theory [44].

The Helmholtz energy of the system can thus be obtained as

$$F = \sum_{k=1}^{m} p^k E^k - TS = \sum_{k=1}^{m} p^k F^k - k_B T \sum_{k=1}^{m} p^k \ln p^k \qquad Eq.\ 2$$

where $E^k$ and $F^k = E^k - TS^k$ are the internal and Helmholtz energies of configuration $k$, respectively. Re-arranging Eq. 2 in the form of partition function, one obtains

$$Z = e^{-\frac{F}{k_B T}} = \sum_{k=1}^{m} e^{-\frac{F^k}{k_B T}} = \sum_{k=1}^{m} Z^k \qquad Eq.\ 3$$

$$p^k = \frac{Z^k}{Z} = e^{-\frac{F^k - F}{k_B T}} \qquad Eq.\ 4$$

where $Z$ and $Z^k$ are the partition functions of the system and configuration $k$, respectively. All above equations reduce to standard statistical mechanics when $S^k = 0$, i.e., pure quantum configurations with $F^k = E^k$. However, the ground-state and stable non-ground-state configurations are not pure quantum configurations, and their Helmholtz energies can be predicted by quasiharmonic approximations [48], which should be accurate in the volume range of experimental observations.

It is noted that Kohn and Sham [4] used the finite temperature generalization of ground-state energy of an interacting inhomogeneous electron gas by Mermin [49] and formulated the entropy of thermal electrons at finite temperatures. Wang et al [48] added the vibrational contribution and presented the Helmholtz energy as follows

$$F^k = E^{k,0} + F^{k,el} + F^{k,vib} = E^k - TS^k \qquad Eq.\ 5$$

$$E^k = E^{k,0} + E^{k,el} + E^{k,vib} \qquad Eq.\ 6$$

$$S^k = S^{k,el} + S^{k,vib} \qquad Eq.\ 7$$

where $F^{k,el}$, $E^{k,el}$, and $S^{k,el}$ are the contributions of thermal electron to Helmholtz energy, internal energy, and entropy of configuration $k$ based on the Fermi–Dirac statistics for electrons, and $F^{k,vib}$, $E^{k,vib}$, and $S^{k,vib}$ are the vibrational contributions to Helmholtz energy, internal energy, and entropy of configuration $k$ based on the Bose–Einstein statistics for phonons, respectively.

With $F^k$ predicted from DFT and $p^k$ from partition functions, the zentropy theory enables the integration of the quantum and statistical mechanics through Eq. 1 to Eq. 7 and has predicted magnetic and ferroelectric phases transitions in a number of materials showing remarkable agreement with experimental observations without models and model parameters [9,41,43,45]. Another important outcome of the zentropy statistical mechanics is the accurate prediction of the free energy landscape including the free energy of unstable states of the system. It can be seen from Eq. 1 to Eq. 7 and was emphasized by Gibbs [36] that the free energy of each configuration is evaluated under the same external constraints as the entire system, i.e., no internal relaxations are considered among configurations, resulting in a macroscopically homogenous system. While this macroscopically homogenous system can be in a stable state, it can also be metastable or unstable with respect to internal processes. This is significant because the common wisdom is that the entropy of an unstable state could not be defined due to its imaginary vibrational modes. However, this static view of a dynamic system considers the vibrations of static particles. While in the zentropy theory, the individual configurations are stable, and it is the statistical competition among configurations that results in the macroscopic instability of the system at its macroscopic critical point or inflection points [50]. The decomposed subsystems from the unstable states all contain the same configurations as macroscopically homogenous system the though with different statistical probabilities in each subsystem. Consequently, the zentropy theory is capable of predicting the entropies of unstable states, thus the free energy barrier between stable states as schematically shown for a system with two stable states in Figure 2 with one internal variable. This naturally leads to the discussion of irreversible thermodynamics in next section.

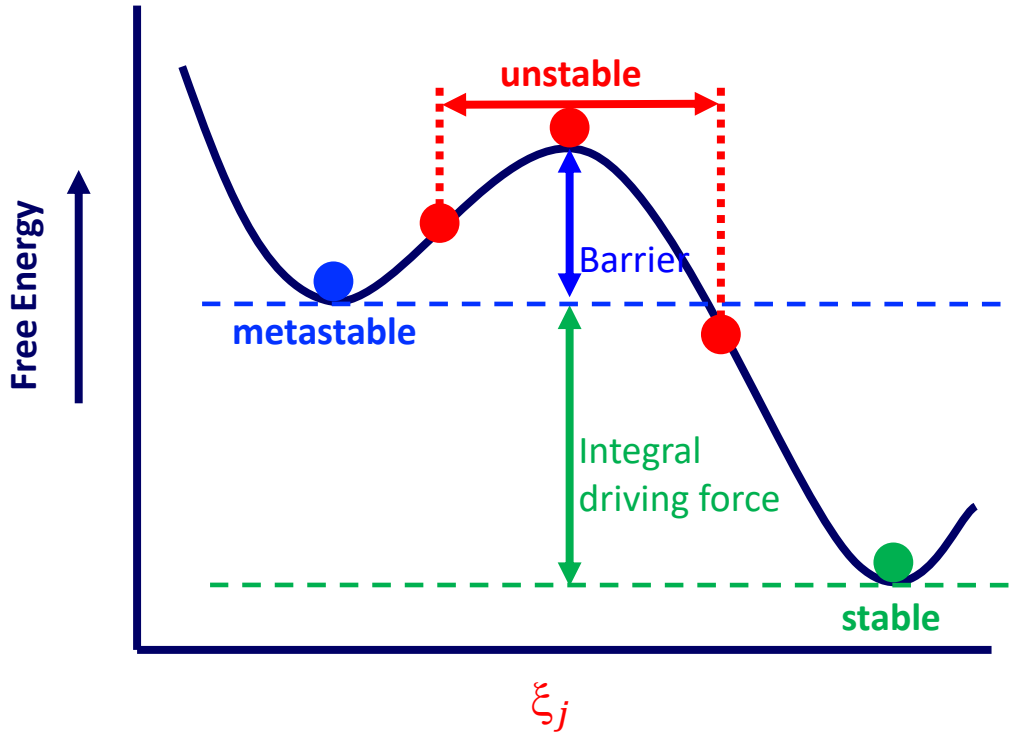

*Figure 2: Schematic diagram of free energy landscape as a function of one internal variable* [41].

## 3 Theory of cross phenomena: Integration of zentropy theory and irreversible thermodynamics

As discussed in details in recent papers by the present author [41,43,44], the combined law of thermodynamics can be written as follows

$$dU = \sum_{a=1}^{n} Y^a dX^a - T d_{ip} S \qquad Eq.\ 8$$

where $Y^a$ and $X^a$ represent the pairs of conjugate variables with $Y^a$ for potentials, such as temperature, stress or pressure, electrical and magnetic fields, and chemical potential, and $X^a$ for molar quantities, such as entropy, strain or volume, electrical and magnetic displacements, and moles of components. $d_{ip}S$ in Eq. 8 denotes the entropy production due to irreversible internal processes ($ip$) inside the system. The first summation in Eq. 8 is to account for the interactions between the system and its surrounding except the entropy change which contains both the exchanges and the internal entropy production, $d_{ip}S$, which is then subtracted by the last term in the equation as discussed in detail in the literature by the present author [41,43,44].

The energy contribution due to the entropy production can be further written as follows

$$Td_{ip}S = \sum_{b=1}^{m} D_b d\xi_b \qquad \text{Eq. 9}$$

$$D_b d\xi_b \geq 0 \qquad \text{Eq. 10}$$

where the summation goes over all independent irreversible internal processes with their driving forces and conjugate molar internal variables denoted by $D_b$ and $\xi_b$, respectively, and Eq. 10 reflects the 2nd law of thermodynamics that an irreversible internal process results in positive entropy production [43,51]. The internal conjugate pairs of $D_b$ and $\xi_b$ can be the same as the external conjugate pairs of $Y^a$ and $X^a$ in Eq. 8 or a combination of some external conjugate pairs such as chemical reactions [52,53] or new sets of variables such as defects in the system. For systems without internal processes, i.e., $D_b d\xi_b = 0$ for every possible internal process, there could be either $D_b = 0$ for system under internal equilibrium with $\xi_b$ becoming a dependent variable or $d\xi_b = 0$ for freeze-in conditions with $\xi_b$ as an independent variable. The combined law derived by Gibbs [54] is with $D_b = 0$ and $dU = \sum_{a=1}^{n} Y^a dX^a$, applicable to equilibrium systems only.

The change rate of $d\xi_b$ is represented by its flux. Based on experimental observations and the top-down view, Onsager postulated phenomenologically a linear relationship of this flux to the gradients of all potentials and further articulated that the coefficient matrix is symmetrical [55,56] with the off-diagonal terms for the cross phenomena, i.e., the flux of a molar quantity by a non-conjugate potential. On the other hand, based on the bottom-up view from the combine law of thermodynamics shown by Eq. 8 without cross terms between non-conjugate variables and the fact that a symmetrical matrix can be diagonalized, the present author postulated that the flux of a molar quantity is proportional to the gradient of its conjugate potential only, i.e., [43,44]

$$J_{\xi_b} = -L_{\xi_b} \nabla D_b \qquad Eq.\ 11$$

where $J_{\xi_b}$ is the flux of the molar internal variable $\xi_b$, and $L_{\xi_b}$ the kinetic coefficient of the internal process.

The cross phenomena are realized through the dependence of $\nabla D_b$ on other independent variables as follows

$$\nabla D_b(\xi_b, \xi_a, D_c) = \frac{\partial D_b}{\partial \xi_b} \nabla \xi_b + \sum_{a \neq b} \frac{\partial D_b}{\partial \xi_a} \nabla \xi_a + \sum_{c \neq a \neq b} \frac{\partial D_b}{\partial D_c} \nabla D_c \qquad Eq.\ 12$$

The corresponding free energy of the system is obtained through Legendre transformation of Eq. 8 as follows

$$d\Phi = d\left(U + \sum_{c \neq b} D_c \xi_c\right) = \sum_{a=1}^{n} Y^a dX^a - \sum_b D_b d\xi_b + \sum_{c \neq b} \xi_c dD_c \qquad Eq.\ 13$$

Based on the Maxwell relation, the last partial derivative in Eq. 12 becomes

$$\frac{\partial D_b}{\partial D_c} = -\frac{\partial^2 \Phi}{\partial D_c \partial \xi_b} = -\frac{\partial \xi_c}{\partial \xi_b} \qquad Eq.\ 14$$

Since each potential is a function of all independent variables in the system, many internal processes take place simultaneously. An internal process stops when all the terms on the right-hand side of Eq. 12 cancel each other so the driving force on the left-hand side of Eq. 12 becomes zero. With Eq. 11 to Eq. 14 representing the irreversible thermodynamics and the free energy $\Phi$ predicted by the zentropy theory discussed in the previous section, a coherent framework for the integration of quantum, statistical, and irreversible thermodynamics is thus established.

All partial derivatives in Eq. 12 are partial 2$^{nd}$ derivatives of free energy as shown by Eq. 14 and represent various properties experimentally measured under equilibrium or freeze-in conditions, i.e., with $D_b d\xi_b = 0$ as discussed above. Those properties are listed in Table 1 and Table 2. There are additional properties related to the 3$^{rd}$ derivatives of free energy such as electro-optic coefficients [57,58].

Table 1: Physical quantities related to the first directives of molar quantities (first column) to potentials (first row), symmetrical due to the Maxwell relations [43,44,59].

|  | $T$, Temperature | $\sigma$, Stress | $E$, Electrical field | $\mathcal{H}$, Magnetic field | $\mu_i$, Chemical potential |
|---|---|---|---|---|---|
| $S$, Entropy | Heat capacity | Piezocaloric effect | Electrocaloric effect | Magnetocaloric effect | $\frac{\partial S}{\partial \mu_k}$ |
| $\varepsilon$, Strain | Thermal expansion | Elastic compliance | Converse piezoelectricity | Piezomagnetic moduli | $\frac{\partial \varepsilon_{ij}}{\partial \mu_k}$ |
| $\theta$, Electrical displacement | Pyroelectric coefficient | Piezoelectric moduli | Permittivity | Magnetoelectric coefficient | $\frac{\partial D_i}{\partial \mu_k}$ |
| $B$, Magnetic induction | Pyromagnetic coefficient | Piezomagnetic moduli | Magnetoelectric coefficient | Permeability | $\frac{\partial B_i}{\partial \mu_k}$ |
| $N_j$, Moles | *Thermoreactivity* | *Stressoreactivity* | *Electroreactivity* | *Magnetoreactivity* | $\frac{\partial N_i}{\partial \mu_k}$, *Thermodynamic factor* |

*Table 2: Cross phenomenon coefficients represented by derivatives between potentials, symmetrical due to the Maxwell relations* [43,44,57].

| | $T$, Temperature | $\sigma$, Stress | $E$, Electrical field | $\mathcal{H}$, Magnetic field | $\mu_i$, Chemical potential |
|---|---|---|---|---|---|
| $T$ | 1 | $-\frac{\partial S}{\partial \varepsilon}$ | $-\frac{\partial S}{\partial \theta}$ | $-\frac{\partial S}{\partial B}$ | $-\frac{\partial S}{\partial c_i}$ Partial entropy |
| $\sigma$ | $\frac{\partial \sigma}{\partial T}$ | 1 | $-\frac{\partial \varepsilon}{\partial \theta}$ | $-\frac{\partial \varepsilon}{\partial B}$ | $-\frac{\partial \varepsilon}{\partial c_i}$ Partial strain |
| $E$ | $\frac{\partial E}{\partial T}$ | $\frac{\partial E}{\partial \sigma}$ | 1 | $-\frac{\partial \theta}{\partial B}$ | $-\frac{\partial \theta}{\partial c_i}$ Partial electrical displacement |
| $\mathcal{H}$ | $\dfrac{\partial \mathcal{H}}{\partial T}$ | $\dfrac{\partial \mathcal{H}}{\partial \sigma}$ | $\dfrac{\partial \mathcal{H}}{\partial E}$ | 1 | $-\dfrac{\partial B}{\partial c_i}$ Partial magnetic induction |
| $\mu_i$ | $\dfrac{\partial \mu_i}{\partial T}$ Thermodiffusion | $\dfrac{\partial \mu_i}{\partial \sigma}$ Stressmigration | $\dfrac{\partial \mu_i}{\partial E}$ Electromigration | $\dfrac{\partial \mu_i}{\partial \mathcal{H}}$ Magnetomigration | $\dfrac{\partial \mu_i}{\partial \mu_j} = -\dfrac{\partial c_j}{\partial c_i} = \dfrac{\Phi_{ii}}{\Phi_{ji}}$ Crossdiffusion |

The present author was first brought into this topic through collaborations on simulations of thermodiffusion where the atomic diffusion is induced by a temperature gradient [60,61], i.e., the Soret effect with the Soret coefficient typically considered as an independent kinetic coefficient based on the Onsager theorem. By the same token, one would expect more than one independent kinetic coefficient for isothermal diffusion in multicomponent systems based on the Onsager theorem. This was puzzling as there is only one diffusion coefficient for each element, i.e., tracer diffusion coefficient or atomic mobility as measured experimentally and modelled computationally [62].

This is particularly interesting in the uphill diffusion in Fe-Si-C alloys where C diffuses from low concentration regions to high concentration regions ($\nabla \xi_b = \nabla c_C$) due to the concentration gradient of Si ($\nabla \xi_a = \nabla c_{Si}$) with a very positive value for $\dfrac{\partial D_b}{\partial \xi_a} = \dfrac{\partial \mu_C}{\partial c_{Si}}$ [43]. Based on Onsager theorem, there would be at least two independent kinetic coefficients for C diffusion, while in the reality, there is only one independent kinetic coefficient related to the tracer diffusivity of C, and the effect of Si are mainly through $\frac{\partial \mu_C}{\partial c_{Si}}$ as shown by Eq. 11 and Eq. 12. This effect is so strong and results in the same sign for $J_{c_C}$ and $\nabla c_C$ in order for C to diffuse from high $\mu_C$ (high $c_{Si}$ and low $c_C$) regions to low $\mu_C$ (low $c_{Si}$ and high $c_C$) regions as shown in the following equation

$$J_{c_C} = -L_{c_C}\nabla \mu_C = -L_{c_C}\left(\frac{\partial \mu_C}{\partial c_C}\nabla c_C + \frac{\partial \mu_C}{\partial c_{Si}}\nabla c_{Si} + \frac{\partial \mu_C}{\partial c_{Fe}}\nabla c_{Fe}\right) \qquad Eq.\ 15$$

The vacancy wind effect occurs when $J_{c_C} + J_{c_{Si}} + J_{c_{Fe}} = -J_{Va}$ with $J_{Va}$ for the flux of vacancy being non-zero [63,64].

The relation between tracer diffusivity and other commonly used diffusivities are shown in Figure 3. Consequently, the thermodiffusion of a component in a multicomponent system can be written as

$$J_{c_b} = -L_{c_b}\nabla \mu_b = -L_{c_b}\left(\frac{\partial \mu_b}{\partial c_b}\nabla c_b + \sum_{a \neq b}\frac{\partial \mu_b}{\partial c_a}\nabla c_a + \frac{\partial \mu_b}{\partial T}\nabla T\right) \qquad Eq.\ 16$$

where the summation includes all components except the diffusion component $b$. The present author's team has applied the same concept to the migration of electrons due to temperature gradient and predicted the Seebeck coefficients for several thermoelectrics as the derivatives of chemical potential of electrons to temperature [65,66], i.e., the last term in Eq. 16 with the diffusion component being electrons. Consequently, the Seebeck coefficient equals to the negative partial entropy of electrons for n-type thermoelectrics and the positive partial entropy of holes for p-type thermoelectrics with details discussed in ref. [43].

Furthermore, the present author's group is working on prediction of $L_{\xi_b}$ based on the free energy barrier between two states of a system predicted by the zentropy theory as shown in Figure 2, without the approximation in the transition state theory used in the DFT-based prediction of self and tracer diffusion coefficients [67–70].

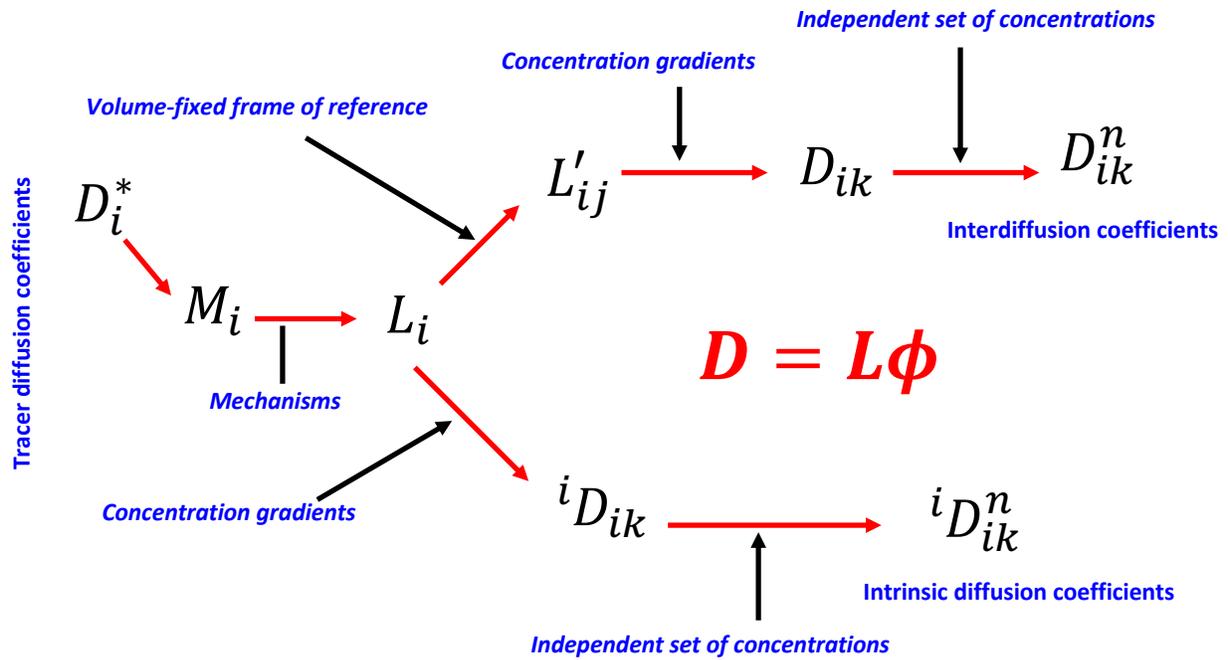

*Figure 3: Relationships among tracer diffusivity, atomic mobility, kinetic L parameters, and intrinsic and chemical diffusivities [43].*

## 4 Summary

In the present paper, a coherent framework for integration of quantum, statistical, and irreversible thermodynamics is discussed. In this framework, the zentropy theory postulates that the entropy of a system contains contributions from entropies of ground-state and symmetry-breaking non-ground-state configurations of the system and the statistical entropy among the configurations. With the free energies of individual configurations predicted from DFT, the zentropy theory integrates quantum and statistical thermodynamics with the partition functions of each configuration calculated from its free energy instead of total energy commonly used in the literature. With accurate free energy of the system predicted by the zentropy theory, theory of cross phenomena postulates that the flux of a molar quantity is proportional only to the gradient of its conjugate potential, and the cross phenomena is due to the dependence of this potential on other non-conjugate molar quantities and potentials, which are related to $2^{nd}$ derivatives of free energy.

**Acknowledgements.** The present review article covers research outcomes supported by multiple funding agencies over multiple years with the most recent ones including the Endowed Dorothy Pate Enright Professorship at the Pennsylvania State University, U.S. Department of Energy (DOE) Grant No. DE-SC0023185, DE-NE0008945, and DE-NE0009288, and U.S. National Science Foundation (NSF) Grant No. NSF-2229690.